\documentclass[aip,reprint]{revtex4-1}

\draft 
\usepackage[pdftex]{graphicx}  

\begin{document}


\title{Impact of Residual Carbon Impurities and Gallium Vacancies on Trapping Effects in AlGaN/GaN MIS-HEMTs} 



\author{Martin Huber}
\affiliation{Infineon Technologies Austria AG, Siemensstrasse 2, A-9500 Villach, Austria}
\affiliation{Johannes Kepler University, Institute of Semiconductor and Solid State Physics, Altenbergerstrasse 69, A-4040 Linz, Austria}

\author{Marco Silvestri}
\affiliation{Infineon Technologies Austria AG, Siemensstrasse 2, A-9500 Villach, Austria}
\author{Lauri Knuuttila}
\affiliation{Infineon Technologies Austria AG, Siemensstrasse 2, A-9500 Villach, Austria}

\author{Gianmauro Pozzovivo}
\affiliation{Infineon Technologies Austria AG, Siemensstrasse 2, A-9500 Villach, Austria}

\author{Andrei Andreev}
\affiliation{Infineon Technologies Austria AG, Siemensstrasse 2, A-9500 Villach, Austria}

\author{Andrey Kadashchuk}
\affiliation{IMEC, Kapeldreef 75, B-3001, Leuven, Belgium}
\affiliation{Institute of Physics, National Academy of Science of Ukraine, Prospekt Nauki 46, 03028 Kyiv, Ukraine}

\author{Alberta Bonanni}
\affiliation{Johannes Kepler University, Institute of Semiconductor and Solid State Physics, Altenbergerstrasse 69, A-4040 Linz, Austria}

\author{Anders Lundskog}
\affiliation{Infineon Technologies Austria AG, Siemensstrasse 2, A-9500 Villach, Austria}

\begin{abstract}
Effects of residual C impurities and Ga vacancies on the dynamic instabilities of AlN/AlGaN/GaN metal insulator semiconductor high electron mobility transistors are investigated. Secondary ion mass spectroscopy, positron annihilation spectroscopy, steady state and time-resolved photoluminescence (PL) measurements have been performed in conjunction with electrical characterization and current transient analyses. The correlation between yellow luminescence  (YL), C- and Ga vacancy concentration is investigated. Time-resolved PL indicating the C$_{\mathrm{N}}$O$_{\mathrm{N}}$ complex as the main source of the  YL, while Ga vacancies or related complexes with C seem not to play a major role. The device dynamic performance is found to be significantly dependent on the C concentration close to the channel of the transistor. Additionally, the magnitude of the YL is found to be in agreement with the threshold voltage shift and with the on-resistance degradation. Trap analysis of the GaN buffer shows an apparent activation energy of $\sim$0.8eV for all samples, pointing to a common dominating trapping process and that the growth parameters affect solely the density of trap centres. It is inferred that the trapping process is likely to be directly related to C based defects.
\end{abstract}

\maketitle 


The AlGaN/GaN material system is a fundamental building-block for the fabrication of high-power GaN on Si metal insulator semiconductor high electron mobility transistors (MIS-HEMTs).~\cite{Kanamura2012, Ueda2014} Despite the large lattice and thermal coefficient mismatch, Si is widely used as the substrate of choice for nitride-based HEMTs. Complex multilayer structures are needed in order to maintain the strain during the growth, at the same time the structure should provide high electrical resistance and low leakage currents. One way to satisfy these requirements, is to introduce acceptor-like impurities, e.g. Fe or C [\onlinecite{Silvestri2013, Poblenz2004}]. However, Fe is considered as non-suitable element in the front-end-of-line Si area, limiting the integration possibilities in low-cost environments. Carbon on the other hand, could potentially fulfill the task as it acts as a source of donor compensation centers, while being Si environment friendly. For efficient GaN based devices, it is of uttermost importance to reduce, or avoid trapping related failures such as current collapse, dynamic on-resistance (R$_{\mathrm{DSON}}$) degradation and threshold voltage (V$_{\mathrm{th}}$) shifts. 

The location of the traps causing dynamic instabilities in the HEMT structure is still under debate. Some studies~\cite{Klein2001a, Verzellesi2014} indicated that trapping occurs solely in the buffer due to C-doping, while others showed that the trapping takes place at the interface between the dielectric and the III-nitride semiconductor.~\cite{Lagger2012, Lagger2014} Combinations of the two previously mentioned processes have also been proposed.~\cite{Meneghesso2014} A clear attribution of the yellow luminescence (YL) band around 2.2 eV in GaN-related materials systems to a single species or defect has been under discussion for long time.~\cite{Reshchikov2005} Previously, it has been argued that Ga vacancies (V$_{\mathrm{Ga}}$) accompanied by C or O could be responsible for the YL band.~\cite{Armitage2003c, Xu2010a} However, according to recent hybrid functional calculations, the V$_{\mathrm{Ga}}$O$_{\mathrm{N}}$ complex is thought to have a maximum in the infrared (IR) region.~\cite{Demchenko2013} Recently, the YL band was attributed to electron transitions from the conduction band minimum to the corresponding transition levels of C substitution at the N site (C$_{\mathrm{N}}$) and at the  C$_{\mathrm{N}}$O$_{\mathrm{N}}$ complex.~\cite{Reshchikov2014} The effect of V$_{\mathrm{Ga}}$ concentration and possible related complexes in the epitaxial structure on dynamic instabilities of working devices has not been thoroughly investigated yet. 

In this work, we report on the impact of buffer residual C impurities and Ga vacancies on the dynamic properties of AlGaN/GaN HEMT structures. We utilize a combination of complementary characterization techniques in order to understand the relationship between YL, impurity concentrations, threshold voltage- and on-resistance instabilities.
\begin{figure*}
\includegraphics[width=1.01\textwidth]{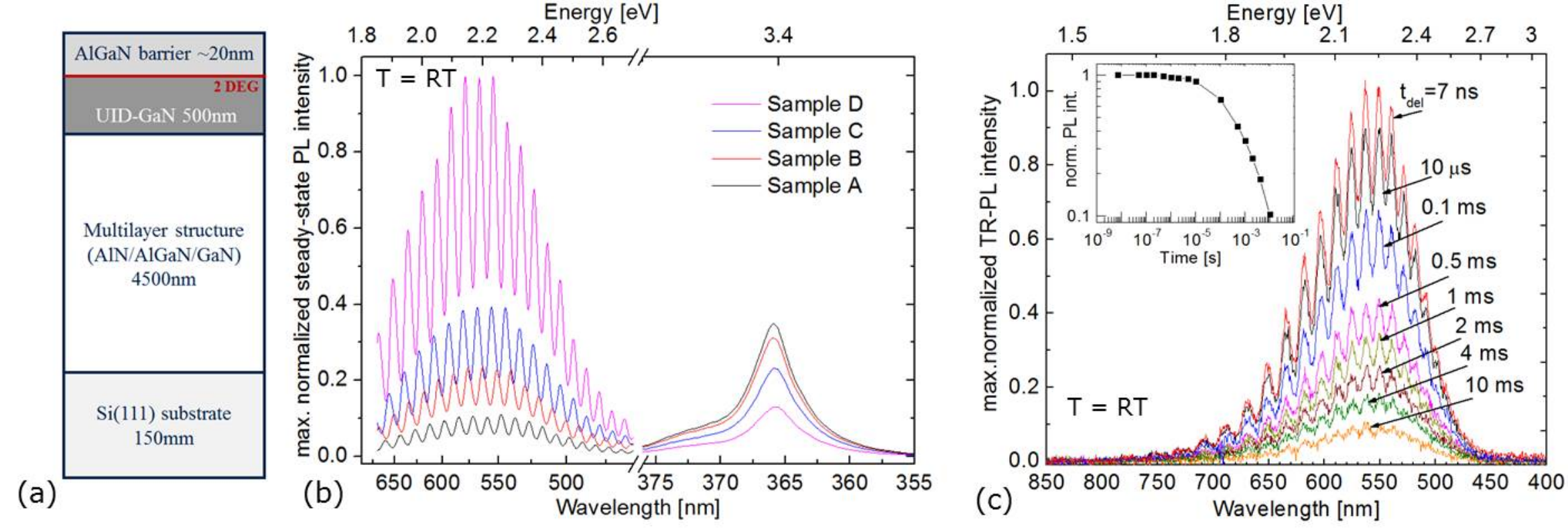}
\caption{(a) Sketch of the studied HEMT structure. (b) Steady-state PL spectra at room temperature showing YL as a broad peak at  $\sim$2.2 eV and the BE peak at  $\sim$3.4 eV. The spectra are normalized to the maximum PL intensity of sample D. (c) TR-PL spectra at room temperature for sample D. Inset: time dependence of the spectrally integrated YL band intensity (squares), the solid line is a guide to the eye.}
\end{figure*}

Four different AlGaN/GaN structures (A, B, C and D) have been grown on Si(111) 150 mm substrates by metal organic vapor phase epitaxy in a multi-wafer AIXTRON planetary reactor. The epitaxial stack is sketched in Fig. 1(a). It consists of three parts, namely a highly resistive multilayer AlGaN/GaN structure grown on the Si substrate, a subsequent unintenionally doped (UID)-GaN layer and an AlGaN barrier overlayer. The thickness of the layers is 4500 nm, 500 nm and 20 nm, respectively. The AlGaN barrier overlayer and the UID-GaN layer form the two dimensional electron gas (2DEG) channel of the HEMT structure. In this study, only the growth conditions of the UID-GaN layer are systematically varied in order to achieve different impurity and vacancy concentrations. Since the variation of the growth conditions has also an effect on the growth rate, the overall UID-GaN layer thickness has been kept constant by adjusting the deposition times.

Secondary ion mass spectroscopy (SIMS) analysis has been performed in a Phi-Evans quadrupole secondary ion mass spectrometry system equipped with Cs primary ion source operated at low range keV energy to determine the impurity concentrations in the UID-GaN layer. 

Room temperature positron annihilation spectroscopy (PAS) is a widely used technique to determine the vacancy concentration in various material systems. For this sample set, PAS analysis has been performed with a slow positron beam using acceleration energies between 1 keV and 25 keV. The $V_G$$_a$ concentration has been determined by analyzing the shape of the Doppler-broadened 511 keV positron-electron annihilation energy peak.~\cite{Tuomisto2013} 

The structures have been characterized by room temperature photoluminescence (PL) measurements using a HeCd laser for excitation with a wavelength of 325 nm and a power density of 0.1 W/mm$^2$. Time-resolved photoluminescence (TR-PL) measurements have been carried out using a set-up consisting of a pulse UV-laser (pulse width of 1.7 ns and pulse energy 300 $\mu$J) for optical excitation at 355 nm and a monochromator coupled to an intensified CCD camera with a time-gated, intensified diode array detector, synchronized to the laser. A variable delay from 7 ns to 10 ms after laser pulse excitation and a constant time gate width of 15 ms have been used to obtain the TR-PL spectra. To increase the signal-to-noise ratio, spectra have been collected by averaging over 200 pulses.

The material characteristics are correlated with the electrical performances of the MIS-HEMT devices fabricated by using a complementary metalo-oxide-semiconductor (CMOS) production line. The 2DEG sheet resistance is $\sim$460 Ohm/sq. Hall mobility and 2DEG density are 1800 cm$^2$/Vs and 7.5x10$^1$$^2$ 1/cm$^2$, respectively. The Au-free ohmic contacts consist of a Ti/Al based metal stack with R$_c$=0.5 Ohm$\cdot$mm. The gate length is 1 $\mu$m, while the gate-source distance and the gate-drain distance are 1.5 $\mu$m and 12 $\mu$m, respectively. The rated breakdown voltage is 650 V at 10 nA/mm. The devices are passivated by using SiN and poly-imide. Electrical characterization includes various stress tests up to 600 V. For the threshold voltage shift, the devices have been subjected to a stepped OFF-state stress, keeping the gate source voltage (V$_{\mathrm{GS}}$ = -12 V) fixed. The stress pulse duration is set to one minute and the drain voltage to 600 V. The V$_{\mathrm{th}}$ and the static R$_{\mathrm{DSON}}$ are measured directly after applying the stress pulse. Double pulse measurements have been performed using an AMCAD double pulse system with a pulse length t$_{\mathrm{on}}$ = 2 $\mu$s and t$_{\mathrm{off}}$ = 1 ms. Dynamic R$_{\mathrm{DSON}}$ are extracted from pulsed-IV curves using different quiescent bias points at V$_{\mathrm{GS}}$ = -20 V and the drain voltage is set to 200V. Transient measurements have been carried out using a parameter analyzer equipped with a heated chuck allowing investigations up to 150 $^{\circ}$C.
\begin{table}
\caption{Impurity- and  V$_{\mathrm{Ga}}$ concentrations in the UID-GaN layer for samples A, B, C and D, respectively. The C and Si content is obtained from SIMS, while the V$_{\mathrm{Ga}}$ concentration from PAS measurements.}
\begin{tabular}{p{1.1cm}p{2.3cm}p{2.3cm}p{2.3cm}}
\hline\hline
Sample&C (10$^1$$^6$/cm$^3$)&Si (10$^1$$^6$/cm$^3$)&V$_{\mathrm{Ga}}$ (10$^1$$^7$/cm$^3$)\\
\hline\hline
A&1.3&5.2&6.0\\
B&3.3&1.0&2.0\\
C&7.0&3.0&1.0\\
D&12.4&1.0&0.8\\
\hline
\end{tabular}
\end{table}

The results of the SIMS and PAS measurements for samples A, B, C and D are summarized in Table I. Significant signals from C and Si are observed while signals from O ($\sim$1x10$^1$$^6$/cm$^3$) and H (low 10$^1$$^7$/cm$^3$) are close to (or below) the SIMS detection limits.~\cite{Evans2007} The variation of growth parameters is known to have a particularly large impact on the V$_{\mathrm{Ga}}$ and C concentrations.~\cite{Armstrong2006} The change in the V$_{\mathrm{Ga}}$ concentration is attributed to the different V/III molar ratios at the different growth conditions.~\cite{Saarinen1998} The samples in Table I are listed in order of increasing C impurity concentration (from A to D).
\begin{figure}
\centering
\includegraphics[scale=1.1]{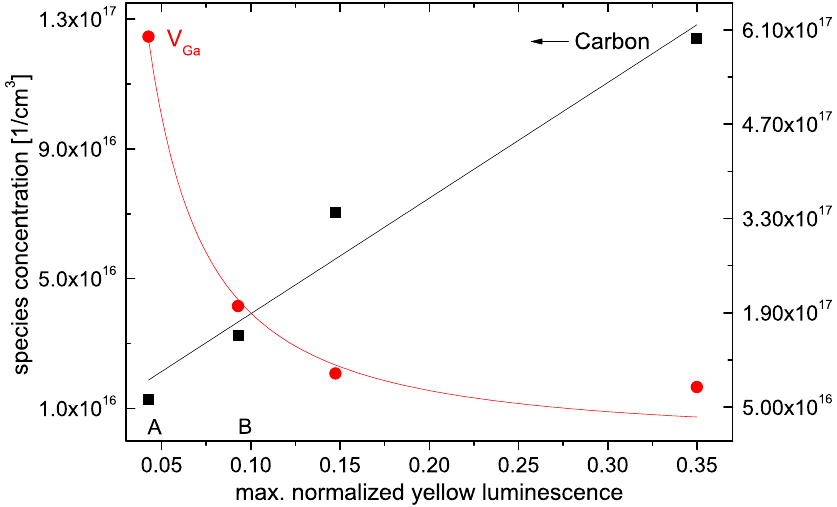}
\caption[FIG. 2]{YL as a function of the C- and V$_{\mathrm{Ga}}$ concentrations. Black squares: SIMS average C concentration in the UID-GaN layer for each sample. Red dots: PAS V$_{\mathrm{Ga}}$ concentration in the UID-GaN layer for each sample.}
\label{FIG. 2}
\end{figure}

In Fig. 1(b) the steady state PL intensity measured at room temperature in the UV-Vis range for the samples A, B, C and D is shown. The PL spectra at room temperature (RT) of all samples are dominated by a broad YL band at around 2.2 eV (564 nm) and by the GaN band-edge (BE) emission peak at 3.4 eV (365 nm). The intensities of both YL and BE band emission are found to vary significantly within the sample set. In order to compare the YL emission from the different samples, peak intensity integration between 1.85 eV and 2.55 eV has been carried out. In Fig. 2 the YL area as a function of the C- and V$_{\mathrm{Ga}}$ concentration in the UID-GaN layers is plotted. Remarkably, a linear correlation between the C concentration and the YL is observed, while the YL band intensity is in anti-correlation with the V$_{\mathrm{Ga}}$-concentration, indicating that the YL is mainly originated by carbon-related defects or by carbon itself in the UID-GaN layer. Indeed, C is generally expected to act as a deep acceptor with the (-/0) charge state transition level at $\sim$0.9 eV above the valance band maximum (VBM) when incorporated at a N site C$_{\mathrm{N}}$ in the GaN lattice.~\cite{Lyons2010} The C$_{\mathrm{N}}$O$_{\mathrm{N}}$ complex is a deep donor with the 0/+ level at $\sim$0.75 eV above the VBM [\onlinecite{Demchenko2013}]. The observed relationship between YL and C concentration is in agreement with a recent comprehensive study by Reshchikov et. al.,~\cite{Reshchikov2014} where it was found that (i) the C$_{\mathrm{N}}$ defect and the C$_{\mathrm{N}}$O$_{\mathrm{N}}$ complex produce the yellow band and that (ii) neither isolated V$_{\mathrm{Ga}}$ nor related complexes can account for this band. Moreover, it was found that the isolated C$_{\mathrm{N}}$ defect does not only produce the YL band (maximum at 2.1 eV) but also a secondary green luminescence (GL) band at 2.4 eV [\onlinecite{Reshchikov2014}], while the C$_{\mathrm{N}}$O$_{\mathrm{N}}$ complex generate solely the YL band with a maximum at 2.2 eV. It is worth mentioning that the GL and YL have distinctly different decay times of a few $\mu$s and a few ms, respectively. The TR-PL measurements presented in Fig. 1(c) do not show presence of the GL band in sample D. This suggests that the observed YL band in our samples is mostly dominated by emission from C$_{\mathrm{N}}$O$_{\mathrm{N}}$ complexes rather than from isolated C$_{\mathrm{N}}$ defects. 
\begin{figure}
\centering
\includegraphics[scale=0.99]{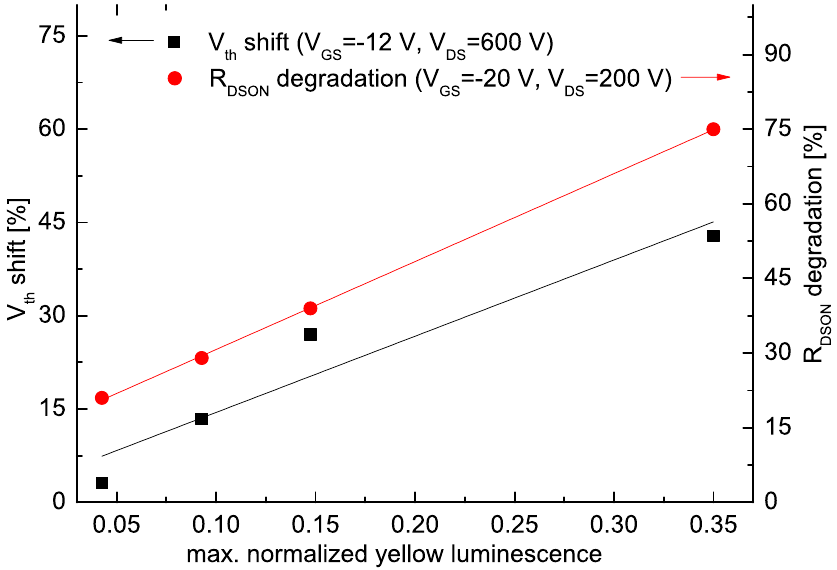}
\caption[FIG. 3]{V$_{\mathrm{th}}$ shift for a 600 V stressed device and R$_{\mathrm{DSON}}$ for a 200 V stressed device measured with a double pulse setup.}
\label{FIG. 3}
\end{figure}
This conclusion is supported by the fact that (i) no short-live component expected for GL in $\mu$s-time domain is actually observed in the measured decay kinetics shown in the inset to Fig. 1(c), and that (ii) the observed YL band has a maximum at 564 nm (2.2 eV) and not at 590 nm (2.1 eV) as it would be expected for C$_{\mathrm{N}}$ defects. 

In order to study the dynamic instabilities of the devices, the threshold voltage shift and the change in R$_{\mathrm{DSON}}$ with the application of a high-voltage off-state stress pulse have been analyzed. Due to trapping processes, V$_{\mathrm{th}}$ is shifted to more negative values and the R$_{\mathrm{DSON}}$ degradation is also changing towards higher values with respect to initial measurement after application of the stress pulses. This behavior was recently attributed to electron traps in the GaN channel that are emptied under measurement conditions and filled under stress conditions.~\cite{Meneghesso2014} Threshhold voltage shift and R$_{\mathrm{DSON}}$ degradation are given in percentage of the initial values.
In Fig. 3 the V$_{\mathrm{th}}$ shift and the R$_{\mathrm{DSON}}$ degradation at one representative stress level for each of the measurements are given as a function of the YL band area. The results reported in Fig. 3 show that the threshhold voltage shift and R$_{\mathrm{DSON}}$ degradation increase with increasing YL band intensity, in agreement with earlier studies.~\cite{Fujimoto2008} We conclude that C-related defects, most likely C$_{\mathrm{N}}$O$_{\mathrm{N}}$ complexes, are the dominant factor for the dynamic instabilities of the final device. Therewith it is confirmed that even at RT the monitoring of the YL is decisive to predict dynamic instabilities and trapping effects in the C compensated HEMTs, if the overall structure and processing sequence are not varied. 
\begin{figure}
\centering
\includegraphics[scale=1]{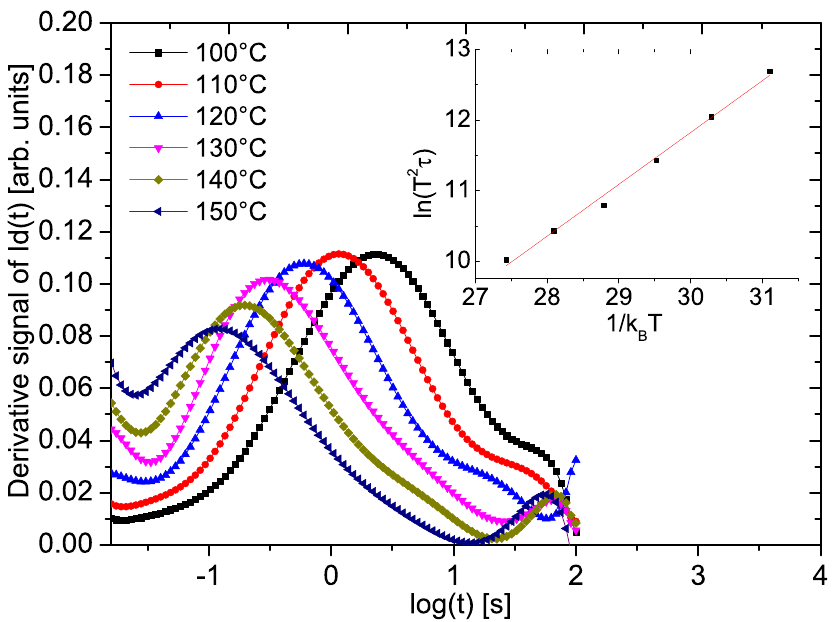}
\caption{Current transient derivatives for sample B at six different temperatures. Inset: Arrhenius plot for sample B.}
\end{figure}

Current transient measurement is a widely used tech-nique to study trap levels in GaN-based HEMTs.~\cite{Joh2011} In order to measure the effect of the multilayer AlN/AlGaN/GaN structure and of the UID-GaN layer solely and to eliminate the effect of surface trapping, an un-gated structure with ohmic contacts and an additional contact on the wafer back-side have been used. With both of the front-side contacts grounded, a negative voltage pulse of -50 V is applied to the wafer back-side for 10 s. In this configuration, the 2DEG acts as an electrode to screen surface influences and causing only buffer traps to be filled.~\cite{Uren2014} Directly after pulsing, the back side voltage is removed and the current between the ohmic contacts on the front-side is recorded keeping one contact at 1 V while the other is grounded. Current transients are then fitted with a polynomial function, as proposed in Ref.  [\onlinecite{Bisi2013}]. The derivatives of the fitted transient signals from sample B are plotted in Fig. 4. The trapping characteristic times are extracted from the peaks of the derivatives and are applied in the Arrhenius plot shown in the inset to Fig. 4. The extracted apparent activation energy and cross section are in the 0.73 eV - 0.83 eV range and mid 10$^-$$^1$$^6$ cm$^2$, respectively. Thorough investigations have been performed to assign defects to the trapping behavior in transient measurements. Particularly the choice of measurement and analysis parameters can influence the apparent activation energies and cross sections that are extracted.~\cite{Bisi2013} However, due to the small variation of the apparent activation energies, we are positive that the dominant trapping mechanism is the same for all the samples and the growth parameters affect the trap concentration solely. In accordance with previous theoretical and experimental results, C-based defects in GaN can be related to the activation energy of $\sim$0.8 eV, taking this value as the energy gap between the deep level and the VBM of GaN [\onlinecite{Lyons2010, Demchenko2013, Uren2014}]. In Fig. 5 the relative transient heights are shown as a function of the C concentration for samples A, B, C and D. These relative transient heights indicate the total amount of charge trapped in the samples. The maximum normalized transients for sample B are shown in the inset to Fig. 5. An incremental trend is observed, indicating that the charge trapping in the devices is related to the C concentration in the UID-GaN layer, in agreement with previous studies.~\cite{Klein2001a}
\begin{figure}
\centering
\includegraphics[scale=0.99]{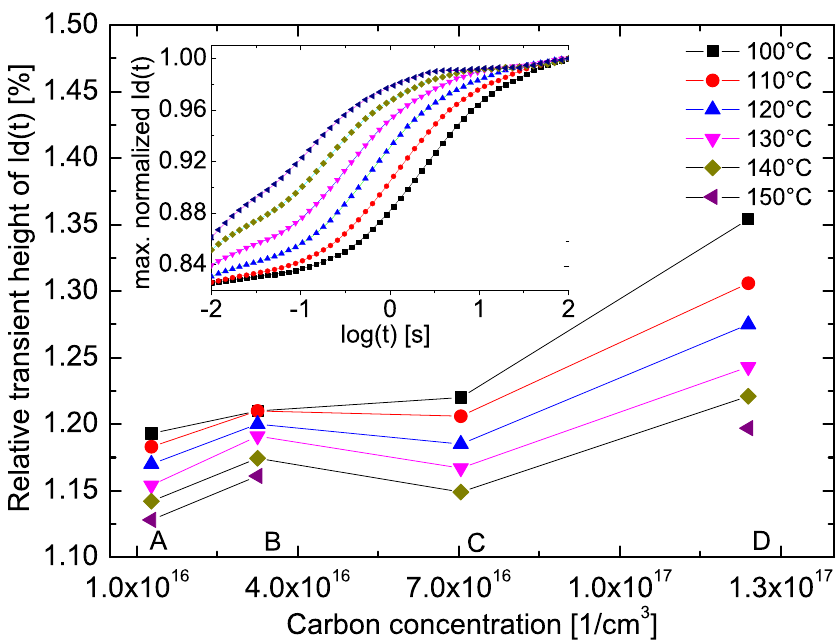}
\caption[FIG. 5]{Relative transient heights as a function of the C concentration.  Inset: normalized transients for sample B.}
\label{FIG. 5}
\end{figure}

In conclusion, dynamic instabilities of four identically processed MIS-HEMT structures with different growth conditions for the UID-GaN layer have been investigated by means of SIMS, PAS, PL and device measurements. The YL intensity is found to have a significant direct correlation with the C concentration and a strong anti-correlation with the V$_{\mathrm{Ga}}$ concentration. 
Moreover,  TR-PL measurements have shown that the observed YL is mostly dominated by emission from C$_{\mathrm{N}}$O$_{\mathrm{N}}$ complexes rather than from isolated C$_{\mathrm{N}}$ defects, other impurities or V$_{\mathrm{Ga}}$-related defects. Furthermore, strong correlations between the YL and  the V$_{\mathrm{th}}$ shift and the R$_{\mathrm{DSON}}$ degradation indicate, that the trapping mechnaism itself is dominantely linked to C-related defects. We conclude that simply room temperature PL measurements can be employed to predict device instabilities already at the level of the unprocessed HEMT epitaxial structure, if the device processing remains unchanged. Our investigations let infer that the dominant trapping process acting in the studied structures has an apparent activation energy of $\sim$0.8eV. Moreover, the results of our transient measurements indicate that the C concentration in the UID-GaN layer close to the channel of the transistors can be employed to manipulate the trapping and the dynamic behavior of the HEMT devices.
\bibliography{library_paper}
\end{document}